\documentclass[twocolumn,prd,showpacs,superscriptaddress,preprintnumbers,nofootinbib]{revtex4}
\usepackage{graphicx}
\usepackage{epsfig}
\usepackage{bm}
\usepackage{latexsym,amssymb,amsmath,amsfonts,amssymb,txfonts,pxfonts,wasysym,float}
\usepackage{color}

\newcommand{\beq}[1]{\begin{equation}\label{#1}}
\newcommand{\eeq}{\end{equation}}
\newcommand{\bea}[1]{\begin{eqnarray} \label{#1}}
\newcommand{\eea}{\end{eqnarray}}
\newcommand{\ba}{\begin{array}}
\newcommand{\ea}{\end{array}}

\def\be{\begin{equation}}
\def\ee{\end{equation}}
\def\gs{\mathrel{
   \rlap{\raise 0.511ex \hbox{$>$}}{\lower 0.511ex \hbox{$\sim$}}}}
\def\ls{\mathrel{
   \rlap{\raise 0.511ex \hbox{$<$}}{\lower 0.511ex \hbox{$\sim$}}}}

\newcommand{\postscript}[2]{\setlength{\epsfxsize}{#2\hsize}
   \centerline{\epsfbox{#1}}}

\begin{document}

\title{Estimating the contribution of Galactic sources to the diffuse neutrino flux}
\author{Luis A.~Anchordoqui}
\affiliation{Department of Physics and Astronomy, Lehman
  College, City University of New York, NY 10468, USA
}

\author{Haim Goldberg}
\affiliation{Department of Physics,
Northeastern University, Boston, MA 02115, USA
}

\author{Thomas \nolinebreak C. \nolinebreak Paul}
\affiliation{Department of Physics and Astronomy, Lehman
  College, City University of New York, NY 10468, USA
}

\affiliation{Department of Physics,
Northeastern University, Boston, MA 02115, USA
}

\author{Luiz~\nolinebreak H.~\nolinebreak M.~\nolinebreak da~\nolinebreak
  Silva}
\affiliation{Department of Physics,
University of Wisconsin-Milwaukee,
 Milwaukee, WI 53201, USA
}

\author{Brian J. Vlcek}
\affiliation{Space and Astroparticle Group, Universidad de Alcal\'a,  Alcal\'a de Henares, E-28871, Spain}

\date{October 2014}
\begin{abstract}
  \noindent Motivated by recent IceCube observations we re-examine the
  idea that microquasars are high energy neutrino emitters. By
  stretching to the maximum the parameters of the Fermi engine we show
  that the nearby high-mass X-ray binary LS 5039 could accelerate
  protons up to above about 20 PeV. These highly relativistic protons
  could subsequently interact with the plasma producing neutrinos up
  to the maximum observed energies. After that we adopt the spatial
  density distribution of high-mass X-ray binaries obtained from the
  deep INTEGRAL Galactic plane survey and we assume LS 5039 typifies
  the microquasar population to demonstrate that these powerful
  compact sources could provide a dominant contribution to the diffuse
  neutrino flux recently observed by IceCube.
\end{abstract}


\maketitle

\section{Introduction}

The IceCube Collaboration has quite recently reported the discovery of
extraterrestrial neutrinos, including 3 events with well-measured
energies around 1~PeV, but notably no events have been observed above
about 2~PeV~\cite{Aartsen:2013bka}. At $E_{
  \nu}=6.3$~PeV, one expects to observe a dramatic increase in the
event rate for $\bar \nu_e$ in ice due to the ``Glashow
resonance'' in which \mbox{${\bar \nu_e}  \, e^- \rightarrow W^- \rightarrow
{\rm shower}$} greatly increases the interaction cross
section~\cite{Glashow:1960zz}. Indeed, the effective detection area
near this resonance becomes about 12 times larger than it is off-peak
value~\cite{Anchordoqui:2004eb}.  However, under the assumption of
democratic flavor ratios, only 1/6 of the total flux is subject to
this enhancement. Integrating the effective area for neutrino
detection from 2 to 10~PeV, we arrive at a factor 40 increase (in the
energy bin centered at the Glashow resonance) compared to the IceCube
sensitivity in the energy bin centered at 1~ PeV. This allows one to
constrain the hypothesis that the neutrino spectrum follows an
unbroken power law.  Under the hypothesis of
an unbroken power law $\propto E_\nu^{-\alpha}$, the effective area between 2 and 10~PeV
together with the 3 observed neutrinos at $\sim 1$~PeV leads an
expectation of a flux which obeys $3\times 40\times 6.3^{-\alpha} \simeq 3
\times 6.3^{2-\alpha}$.  For zero events observed (and none expected
from background), Poisson statistics implies that fluxes predicting
more than 1.29 events are outside the 68.27\%
CL~\cite{Feldman:1997qc}.  Consistency within 1~$\sigma$ then requires
$\alpha \geq 2.5$ for energies above about 2~PeV. The event rate
derived ``professionally''~\cite{Barger:2014iua} differs by a tiny
factor from our back-of-the-envelope estimate. If we assume canonical
Fermi shock acceleration dominates below this energy, we would then
require a break with a magnitude of roughly $\Delta \alpha = 0.5$.

We note in passing that the strong suppression observed in the ultra
high energy cosmic ray (UHECR) spectrum ($\propto E^{-\gamma}$) at
$E\sim 40$~EeV corresponds to a spectral index change from $\gamma
\sim 2.6$ to $\gamma \sim 4.3$, or $\Delta \gamma \sim
1.7$~\cite{Abraham:2010mj}.  This suppression may be due to
interactions of UHECRs {\it en route} to Earth, or it may represent a
natural acceleration endpoint. Indeed, composition data from the
Pierre Auger Observatory tend to favor the latter scenario, or
possibly a combination of the two effects~\cite{Anchordoqui:2013eqa}.
If the strong UHECR spectrum does indeed reflect an acceleration
endpoint, it appears that the smaller cutoff of the energy spectrum
for neutrinos  could also plausibly be
attributed to such an effect. Hereafter we assume the spectral break
does in fact represent an acceleration end point~\cite{Anchordoqui:2014hua}.

Given the overall isotropy of the observed $\nu$ arrival directions
and the fact that one of the 3 highest energy events arrives from
outside the Galactic plane, one might suspect an extragalactic origin
for the extraterrestrial neutrinos.  If the neutrino sources are
extragalactic, the $\gamma$-rays expected to accompany the $\nu$'s
saturate the $\gamma$ flux observed by the Fermi satellite for a
neutrino spectrum with $\alpha \approx 2.15$~\cite{Murase:2013rfa}.
The statistical analysis sketched above, taken together with the
constraint on the spectral index derived from Fermi measurements,
points to a spectral cutoff, which precludes a rate increase near the
Glashow resonance.

Several explanations have been proposed to explain the origin of
IceCube's events~\cite{Anchordoqui:2013dnh}. Interestingly, a priori
predictions for the diffuse $\nu$ flux from FRI
radiogalaxies~\cite{Anchordoqui:2004eu} and
starbursts~\cite{Loeb:2006tw} provide a suitable $\alpha$ and
normalization for the $\nu$ flux while simultaneously retaining
consistency with a cutoff at $E_\nu \sim 3~{\rm
  PeV}$~\cite{Anchordoqui:2014yva}. Other potential sources that can
partially accommodate IceCube data include gamma-ray
bursts~\cite{Waxman:1997ti}, clusters of
galaxies~\cite{Murase:2008yt} (see however~\cite{Zandanel:2014pva}),
and active galactic nuclei~\cite{Kalashev:2013vba}.  However, the identification of
extragalactic neutrino point-sources from a quasi-diffuse flux is
challenging due to the (large) atmospheric neutrino
background~\cite{Ahlers:2014ioa}.

\begin{figure}[tbp]
\postscript{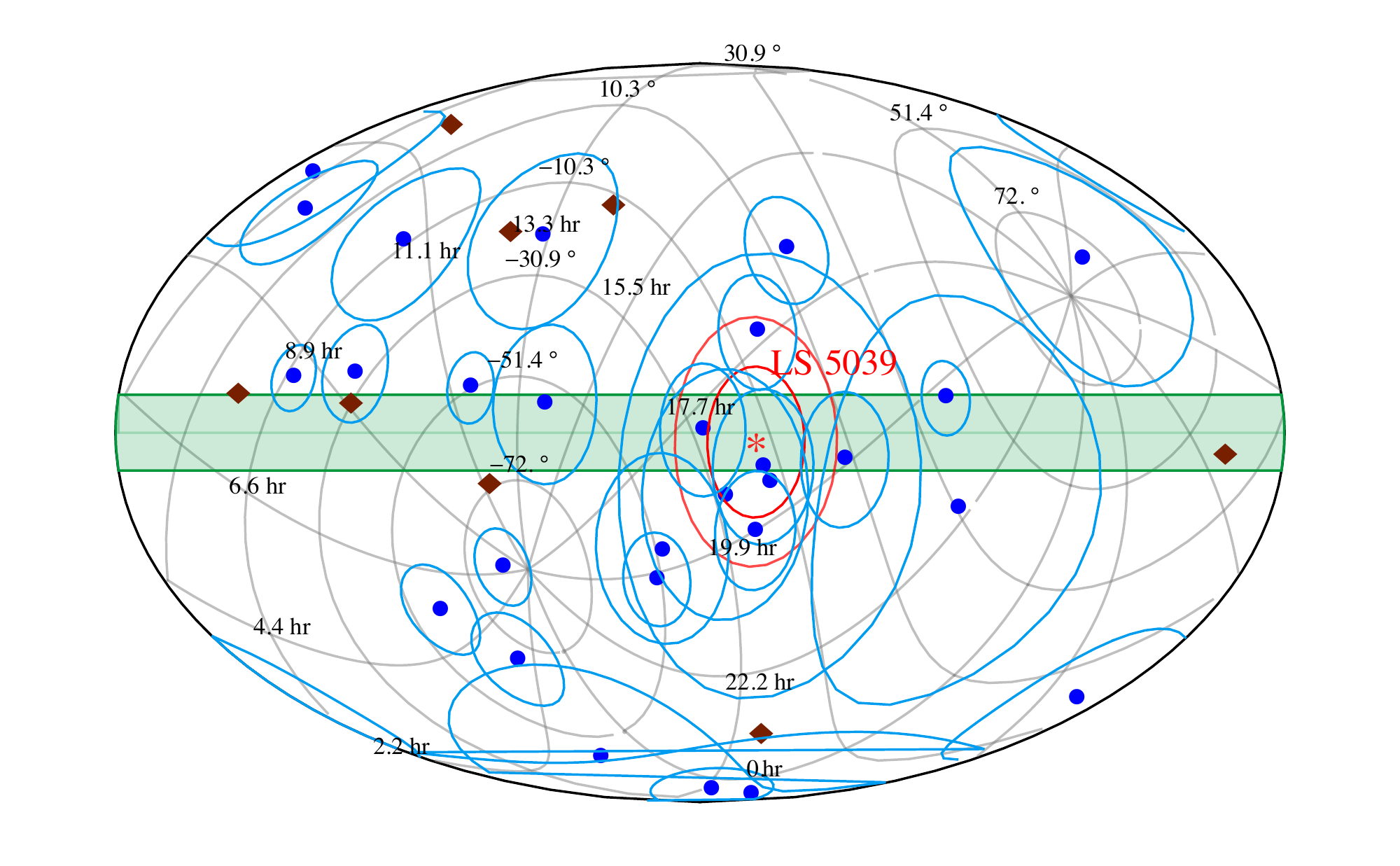}{0.99}
\caption{The 27 shower events (circles) and 8 track events
  (diamonds) reported by the IceCube Collaboration in equatorial
  coordinates. The asterisk indicate the location of LS 5039 and
  the circular contours centered at this position correspond to radii
  of $15^\circ$ and $25^\circ$.  These contours are designed to help the
  reader understand how much weight each point contributes to
  likelihood. The shaded band delimits the Galactic
  plane.}
\label{fig:uno}
\end{figure}

On the basis of existing data a significant contribution from Galactic
sources cannot yet be excluded~\cite{Fox:2013oza,Anchordoqui:2013qsi}.
Searches for multiple correlations with the Galactic plane have been
recently reported by the IceCube
Collaboration~\cite{Aartsen:2013bka}. When letting the width of the
plane float freely, the best fit corresponds to $\pm7.5^\circ$ with a
post-trial chance probability of 2.8\%, while a fixed width of
$\pm2.5^\circ$ yields a p-value of 24\%. In particular, some of the
events seem to cluster near the Galactic
center~\cite{Razzaque:2013uoa}, which has been whimsically described
as a neutrino lighthouse~\cite{Bai:2014kba}. Indeed, a particularly
compelling source of some of these neutrinos could be LS
5039~\cite{Aartsen:2013bka}.  Figure~\ref{fig:uno} contains a display
of the shower and track events reported by the IceCube
Collaboration~\cite{Aartsen:2013bka}. Using these data, the
Collaboration conducted a point source search using an un-binned
maximum likelihood method described in~\cite{Braun:2008bg}.  For both
the clustering and point source search, the number of estimated signal
events, $x_s$, is left as a free parameter and the maximum of the
likelihood is found at each location.  For the point source search,
the most significant source is the binary system LS 5039, with a value
of $x_s = 4.9$, and a corresponding p-value of 0.002.  Of course there
are many sources in the sky; whether this one turns out to be a good
candidate, time will tell.

In summary, though the clustering is not statistically significant one
cannot rule out a Galactic origin for some of these events. Motivated by this
fact we perform a generalized calculation of the flux expected from
various source distributions, taking account of the location of the
Earth in the Galaxy.  In particular, we reduce the problem to two
specific parameters, the distance to the nearest source and the
overall population density. LS 5039 has been discussed in the
literature as potential high energy neutrino
emitter~\cite{Aharonian:2005cx}.  We consider this source as specific
example and assume it typifies the population of Galactic
microquasars ($\mu$QSOs).\footnote{$\mu$QSOs are a sub-class of X-ray binary
  systems that produce collimated outflows observed as non-thermal
  radio structures~\cite{Mirabel:1994rb}. This particular morphology
  probably originates in relativistic jets launched from the inner
  parts of accretion disks around stellar mass black holes or neutron
  stars~\cite{Mirabel:1999fy}.} We generalized the argument such that
it can be applied to various source populations. First we bracket the
realm of plausibility and consider a uniform distribution and an
exponential distribution peaked at the Galactic center. For
illustrative purposes, we consider several conceivable different
distances to the nearest source. After that we turn our attention to
the interesting possibility of $\mu$QSOs for which the overall
distribution of surface density in the Galaxy has a peak at
galactocentric radii $5 - 8~{\rm kpc}$~\cite{Lutovinov:2013ga,Grimm:2001vd}.

The layout of the paper is as follows. In Sec.~\ref{sec-dos} we
revisit the model presented in~\cite{Aharonian:2005cx} in order to
better estimate the expected neutrino flux, especially in the PeV
region. In Sec.~\ref{dos-extra} we compare the properties of LS 5039
with other Galactic microquasars, showing that LS 5039 provides a
reasonable lower bound on the power of this type of source.  In
Sec.~\ref{sec-tres} we estimate the contribution of Galactic sources
to the overall diffuse neutrino flux on the assumption that LS 5039
typifies the population. By comparing this estimate with IceCube data
we find the minimum neutrino production efficiency required to
dominate the spectrum. In Sec.~\ref{cuatro-extra} we employ
constraints from $\gamma$-ray observations to bolster our
hypothesis. We also address the relevance of our previous
finding~\cite{Anchordoqui:2013qsi} that a spectral index of 2.3 is
consistent with the most recent IceCube spectral shape as well as
current bounds on cosmic ray anisotropy.  Our conclusions are
collected in Sec.~\ref{sec-cuatro}.

\section{IceCube neutrinos as the smoking ice of LS 5039 engine}
\label{sec-dos}

LS 5039 is a high-mass X-ray binary (HMXB) system that displays
non-thermal persistent and variable emission from radio frequencies to
high-energy (HE), $E_\gamma > 100~{\rm MeV}$, and very-high-energy
(VHE), $E_\gamma > 100~{\rm GeV}$, gamma rays. The system contains a
bright ON6.5 V((f)) star~\cite{Clark,McSwain:2003ei} and a compact
object of unknown nature. This degenerate companion has a mass between
1.4 and 5 $M_\odot$~\cite{Casares:2005ig}. The orbit of the system has
a period of 3.9 days and an eccentricity around
0.35~\cite{Casares:2005ig,Aragona:2009rx,Sarty:2010gg}.  The distance
to the source has recently been updated to $2.9 \pm 0.8~{\rm
  kpc}$~\cite{Moldon:2012xk}. At the apastron the orbital separation
of the binary system is $2.9 \times 10^{12}~{\rm cm}$ and becomes $1.4
\times 10^{12}~{\rm cm}$ at periastron~\cite{Casares:2005ig}.
Variability consistent with the orbital period in the energy range
$100~{\rm MeV} \alt E_\gamma \alt 300~{\rm GeV}$ was detected by
Fermi~\cite{Abdo:2009sg}.  The system is also a TeV emitter, with
persistent, variable, and periodic emission, as detected by
H.E.S.S.~\cite{Aharonian:2005eb,Aharonian:2006vu}. The overall luminosity in the frequency band ${\rm keV} \alt E_\gamma \alt
{\rm GeV}$ is $L \sim 10^{35}~{\rm erg} \, {\rm
  s}^{-1}$~\cite{Paredes:2001nu}.

Whether the HE/VHE gamma rays are a of hadronic or leptonic origin is
a key issue related to the origin of Galactic cosmic rays. In all
gamma-ray binaries, the nature of the compact object is fundamental
for understanding the physical processes involved in the particle
acceleration that is responsible for the multi-wavelength emission. If
the compact object is a black hole, the accelerated particles would be
powered by accretion, and produced in the jets of a $\mu$QSO. On
the other hand, if the compact object is a young non-accreting pulsar,
the particle acceleration would be produced in the shock between the
relativistic wind of the pulsar and the stellar wind of the massive
companion star. The detection of elongated asymmetric emission in
high-resolution radio images was interpreted as mildly relativistic
ejections from a $\mu$QSO jet and prompted its identification with
an EGRET gamma-ray
source~\cite{Paredes:2001nu,Paredes:2002vy}. However, recent Very Long
Baseline Array observations~\cite{Ribo:2008pd} show morphological
changes on short timescales that might be consistent with a pulsar
binary
scenario~\cite{Dubus:2006ze,SierpowskaBartosik:2008av,SierpowskaBartosik:2008yp}. On
the other hand, no short-period pulsations were observed either in
radio~\cite{McSwain:2011cd} or X-rays~\cite{Rea:2011kq} definitively
demonstrating the compact object to be a pulsar.  New IceCube data
will clarify this situation, as the only plausible high energy
neutrino emission mechanism requires a compact object powering jets.

\begin{figure}[tbp]
\postscript{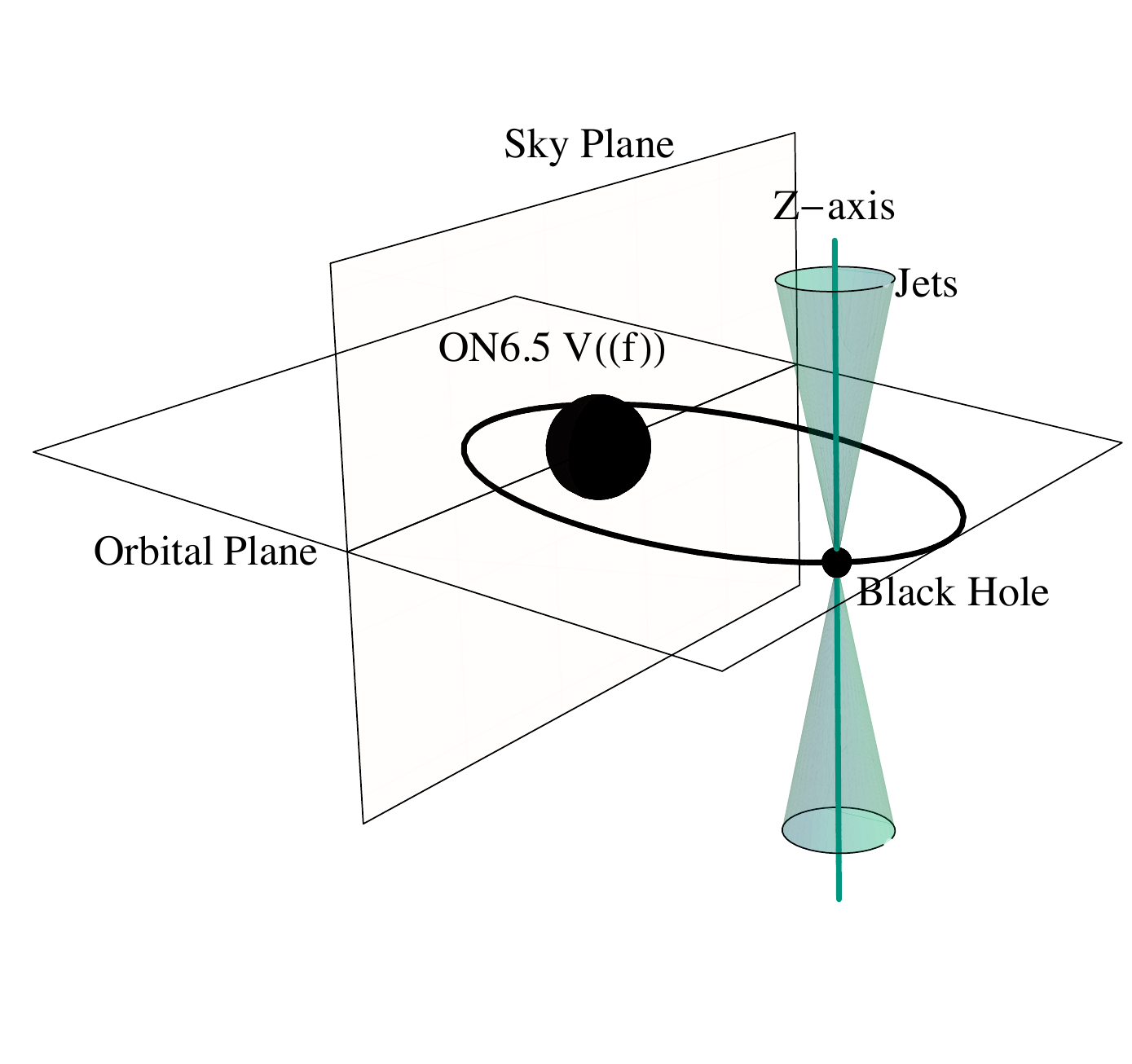}{0.99}
\caption{Sketch of the binary system.}
\label{fig:dos}
\end{figure}

Simultaneous production of $\gamma$'s and $\nu$'s generally requires
two components: {\it (i)} an effective proton accelerator, up to $E
\approx 16 \, E_{\nu}^{\rm max}$ and beyond; {\it (ii)} an effective
target (converter). The maximum observed neutrino energies then
require proton acceleration up to at least $E \agt 20$~PeV. The most
likely site for particle acceleration in LS 5039 is the jet, which
with a speed $v = 0.2 c$ and a half-opening angle $\theta \alt
6^\circ$ extends out to 300  milliarcsecond (mas), that is about $ 10^{16}~{\rm
  cm}$~\cite{Paredes:2002vy}.  Within the inner parts of the jet,
with a radius $R_{\rm jet} \sim 10^9$~cm, a magnetic field $B \agt
10^5~{\rm G}$ could be sufficient to boost protons up to very high
energies.  The maximum proton energy is determined by the Hillas
condition $r_L \leq R_{\rm jet}$, which gives
\begin{equation}
E_{\rm max} \alt 30 \, \left(\frac{R_{\rm jet}}{10^9~{\rm
  cm}}\right) \, \left(\frac{B}{10^5~{\rm G}}\right)~{\rm PeV} \,,
\end{equation}
where $r_L$ is the Larmor radius. A value compatible with this maximum
energy has been obtained in an independent
calculation~\cite{Levinson:2001as}. The accelerated protons can
interact efficiently with the ambient cold plasma throughout the
entire jet. In what follows we assume that the base of the jet is
located close to the inner parts of the accretion disk, that is, the
jet axis $z$ is taken normal to the orbital plane, as shown in
Fig.~\ref{fig:dos}. Here, $z_0 \sim 30 R_{\rm S}$, where
\begin{equation}
R_{\rm S} \simeq 3 \times 10^5 \, \left(\frac{M_{\rm BH}}{M_\odot}\right)~{\rm cm}
\end{equation}
is the Schwarzschild radius. If the magnetic field drops as $B \propto
z^{-1}$, the condition of the confinement of protons in the jet,
$r_{\rm L} \leq R$ implies $E_{\rm max} \propto B z$=constant, where
$R=\theta z$ is the radius of the jet at a distance $z$. Thus, one may
expect acceleration of protons to the same maximum energy $E_{\rm
  max}$ over the entire jet region. However, if there is a faster drop
of $B$ with $z$, the protons at some distance $z_t$ from the compact
object will start escaping the jet. If this happens within the binary
system, i.e.  $z_t \leq 10^{12} \rm cm$, protons interacting with the
dense wind of the optical star will result in additional $\gamma$-ray
and neutrino production outside the jet.

If the jet power is dominated by the kinetic energy of bulk motion of
cold plasma, the baryon density of the jet $n_{\rm jet}$ can be
estimated from the jet power,
\begin{equation}
L_{\rm jet}=\frac{\pi}{2} \, R_{\rm jet}^2(z)
\,n_{\rm jet}(z)\, m_p v^3 \, .
\end{equation}
The efficiency of $\gamma$-ray
production in the jet is
\begin{equation}
\rho_\gamma= \frac{L_\gamma}{L_p} = \sigma_{pp} f_\pi
\int_{z_0}^{z_t} n_{\rm jet}(z) dz \leq 1 \,,
\end{equation}
where $L_\gamma$ is the luminosity of VHE $\gamma$-rays
and $L_p$ is the  power of accelerated protons. Here,
$\sigma_{pp} \approx 40$~mb is
the cross-section of inelastic  $pp$ interactions,
and $f_\pi \approx 0.15$ is the fraction of the
energy of the parent proton transfered to a
high energy $\gamma$-ray~\cite{Frichter:1997wh}. Given the recent estimate
of the black hole mass in LS~5039
$M=3.7_{-1.0}^{+1.3} \, M_\odot$~\cite{Casares:2005ig}, we set
$z_0 \simeq 3 \times 10^7 \ \rm cm$.
For the profile of the number density,  we adopt a power law form
$n_{\rm jet} = n_0 (z_0/z)^{-s}$, where $s=0$ for a cylindrical
geometry, $s=2$ for a conical jet, and $s=1$ for the intermediate case.
Expressing the acceleration power of protons
in terms of the total jet power,
$L_p=\kappa L_{\rm jet}$,
one  finds the following requirement for the jet power,
 \begin{equation}
 L_{\rm jet} \approx 2 \times 10^{37}\,
\frac{L_{\gamma,34}^{1/2} (v/0.2c)^{3/2}}
 {\sqrt{{\cal C}(s) \kappa/0.1}}
 \ \rm erg \ s^{-1} \ ,
 \label{jetpower}
 \end{equation}
 where $L_{\gamma,34}= L_{\gamma}/10^{34} \ \rm erg \ s^{-1}$ and
 $\kappa$ is the acceleration efficiency.  The parameter ${\cal C}(s)$
 characterizes the geometry/density profile of the jet: for $s=0, \ 1,
 \ 2$, we find ${\cal C}(s)=z_t/z_0, \, \ln(z_t/z_0),$ and 1,
 respectively.  The  cylindrical geometry provides the highest
 efficiency of $\gamma$-ray production. However, since $L_\gamma \alt
 1/30 L_{\rm jet}$ (assuming $\approx 10\%$ efficiency of proton
 acceleration, and taking into account that the fraction of energy of
 protons converted to $\gamma$-rays cannot exceed 30\%) the
 $\gamma$-ray production cannot be extended beyond $z_t \sim 10^4 z_0
 \sim 3 \times 10^{11} \rm cm$.  The conical geometry corresponds
 to the minimum efficiency of $\gamma$-ray production, and thus the
 largest kinetic power of the jet. In this case the bulk of
 $\gamma$-rays are produced not far from the base.  For $s=1$,  $\gamma$-rays are produced in equal amounts
 per decade of length of the jet, until the jet terminates.

If $\gamma$-rays are indeed produced in $pp$ interactions, one would
expect production of high energy neutrinos at a rate close to the
$\gamma$-ray production rate. However, since $\gamma$-rays are subject
to energy-dependent absorption, both the energy
spectrum and the absolute flux of neutrinos,
\begin{equation}
\phi_\nu (E_\nu) \simeq 2 \ \phi_\gamma(E_\gamma)\ \exp[\tau(E_\gamma)],
\end{equation}
could be quite different from that of the detected $\gamma$-rays, where $E_\nu
\simeq E_\gamma/2$. The optical
depth $\tau (E)$ depends significantly on the location of the
$\gamma$-ray production region, and therefore varies with time if this
region occupies a small volume of the binary system.  This may lead to
time modulation of the energy spectrum and the absolute flux of TeV
radiation with the orbital period~\cite{Bottcher:2005pj}.
Moreover, the $\gamma \gamma$ interactions generally cannot be reduced
to a simple effect of absorption. In fact, these interactions initiate
high energy electron-photon cascades, driven by  inverse Compton
scattering and $\gamma \gamma$ pair production. The cascades
significantly increase the transparency of the source.  The spectra of
$\gamma$-rays formed during the cascade development significantly
differ from the spectrum of $\gamma$-rays that suffer only absorption.

\begin{figure}[tbp]
\postscript{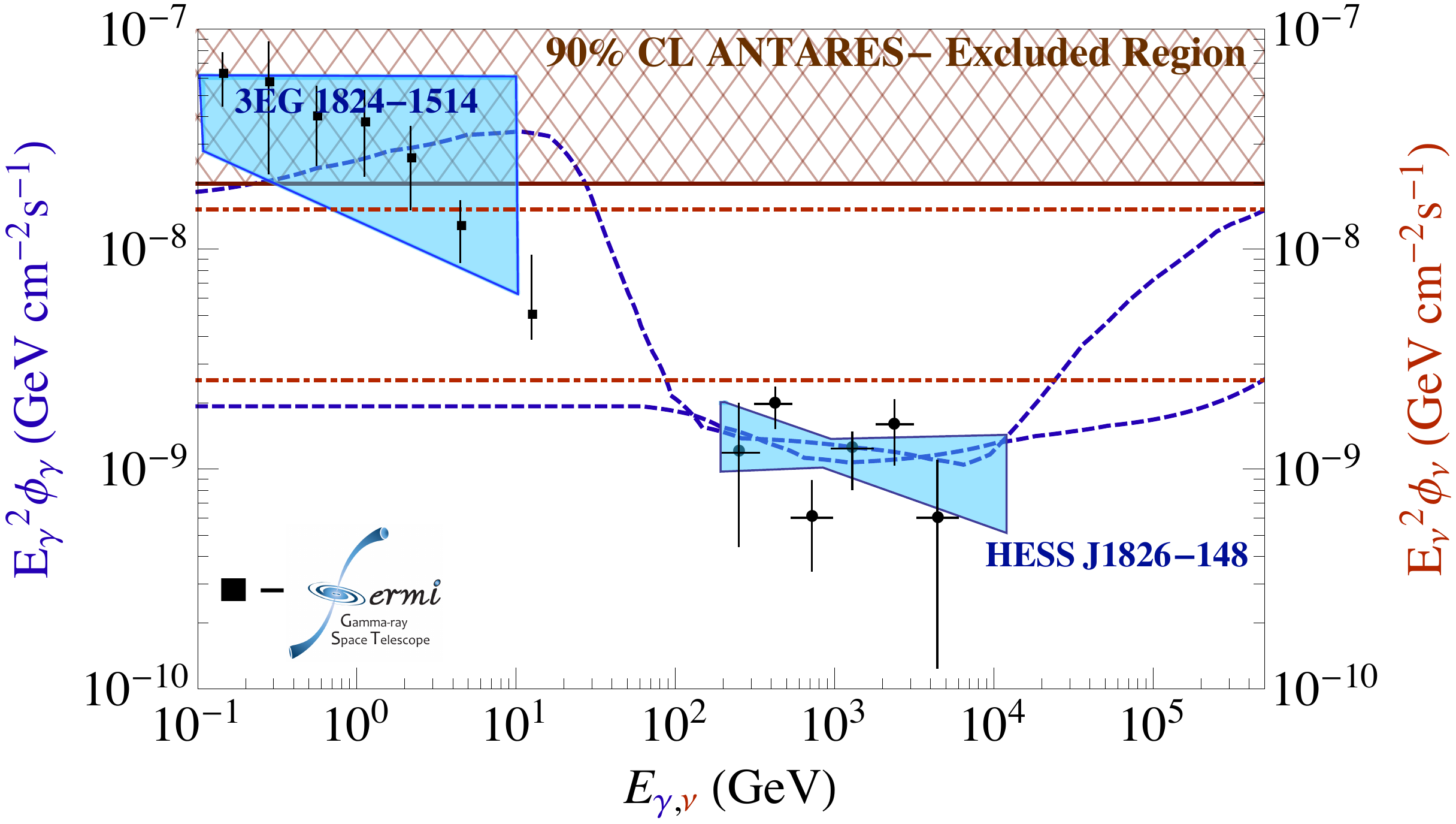}{0.99}
\caption{The dashed curves represent the time averaged $\gamma$-ray
  spectra of LS 5039 after cascading in the anisotropic radiation
  field of the normal companion star. The curves are normalized to
  reproduce the observed $\gamma$-ray flux by H.E.S.S. in the TeV
  range~\cite{Aharonian:2005eb,Aharonian:2006vu}. If pions are
  produced near the base of the jet, the $\gamma$'s produced through
  $\pi^0$ decay can trigger cascades in the plasma, yielding a photon
  flux which can marginally accommodate EGRET~\cite{Hartman:1999fc}
  and Fermi~\cite{Abdo:2009sg} data.  The dot-dashed horizontal lines
  indicate the accompanying neutrino flux.  All curves are averaged
  over the orbital period taking into account data on the geometry of
  the binary system~\cite{Casares:2005ig}. The cross-hatched area
  indicates the 90\% upper limit on the flux from LS 5039 reported by
  the ANTARES Collaboration~\cite{Adrian-Martinez:2014wzf}.  }
\label{fig:tres}
\end{figure}

To model the electromagnetic cascade developed in the plasma we adopt
the method described in~\cite{Aharonian:2002fe}. In our calculations
we include the three dominant processes driving the cooling of the
electromagnetic cascade: photon-photon pair production, inverse
Compton scattering, and synchrotron radiation from electrons. Because of
the orbital motion, both the absolute density and the angular
distribution of the thermal radiation of the star relative to the
position of the compact object  vary with time. We take
into account the effect induced by the anisotropic (time-dependent)
distribution of the target photons on the Compton scattering and
pair-production processes~\cite{Khangulyan:2005ff}.  We normalize the
cascade spectrum of photons to the flux reported by the H.E.S.S.
Collaboration in the TeV energy
range~\cite{Aharonian:2005eb,Aharonian:2006vu}. Interestingly, if pion
production is mostly dominated by collisions close to the base of the
jet (i.e. $z \alt 10^{8}~{\rm cm}$) then the resulting flux of
$\gamma$-rays can marginally accommodate observations in the
GeV-range~\cite{Abdo:2009sg,Hartman:1999fc}. However, if pion
production takes place well above the base of the jet ($z =
10^{13}~{\rm cm}$) the flux of GeV-photons  becomes about
an order of magnitude smaller. These two extreme situations, which are shown in
Fig.~\ref{fig:tres}, provide an upper and a lower bound on the
resulting neutrino flux
\begin{equation}
\phi_\nu (E_\nu) = \zeta \, \ E_\nu^{-2}~{\rm GeV}^{-1} \ {\rm cm}^{-2} \ {\rm s}^{-1} \, ,
\label{fluxls}
\end{equation}
where $1.8 \times 10^{-9} < \zeta < 1.6 \times 10^{-8}$. The lower
value of $\zeta$ is in good agreement with the results of
Ref.~\cite{Distefano:2002qw}.\footnote{The two analyses  assume the
  same fiducial value for $\kappa$. Good agreement is achieved by taking the fiducial value for the fraction of the jet kinetic
energy which is converted to internal energy of
electrons and magnetic fields.} It is notable that while our results
are ultimately derived from demanding consistency between neutrino and
photon data, the results in Ref. ~\cite{Distefano:2002qw} are derived
from assumption on source parameters. For a source distance $d
\simeq 3~{\rm kpc}$, the flux range given in (\ref{fluxls})
corresponds to an integrated luminosity per decade of energy,
\begin{eqnarray}
L_\nu^{^{LS\, 5039}} & = & 4 \pi d^2  \int_{E_1}^{E_2}   E_\nu \,
\phi(E_\nu) \, dE_\nu \nonumber \\ & = & 4 \pi \, \left(\frac{d}{{\rm
      cm}}\right)^2 \, \zeta \,
 \ln 10~{\rm GeV}  \  {\rm s}^{-1},
\end{eqnarray}
 in the range $7.0 \times
10^{33}~{\rm erg} \, {\rm s}^{-1}  \alt L_\nu^{^{LS \, 5039}} \alt 6.4 \times
10^{34}~{\rm erg} \, {\rm s}^{-1}$.

Herein we have assumed the usual Fermi injection spectral index of
$\alpha =2$. The spectral index of $\gamma$-radiation measured by
H.E.S.S. varies depending upon the orbital configuration, reaching a
maximum value of 2.53~\cite{Aharonian:2005eb,Aharonian:2006vu}. In the
next two sections we will assume the ``traditional'' spectral
index. In Sec.~\ref{cuatro-extra} we comment on the effect of a steeper
spectrum.

Determining whether this analysis can be straightforwardly generalized
to all sources in the Galaxy depends on whether neutrino emission from
LS 5039 can typify the population of $\mu$QSOs. It is this that we
now turn to study.

\section{Generalities of the microquasar population in the Galaxy}
\label{dos-extra}

\begin{figure}[tbp]
\postscript{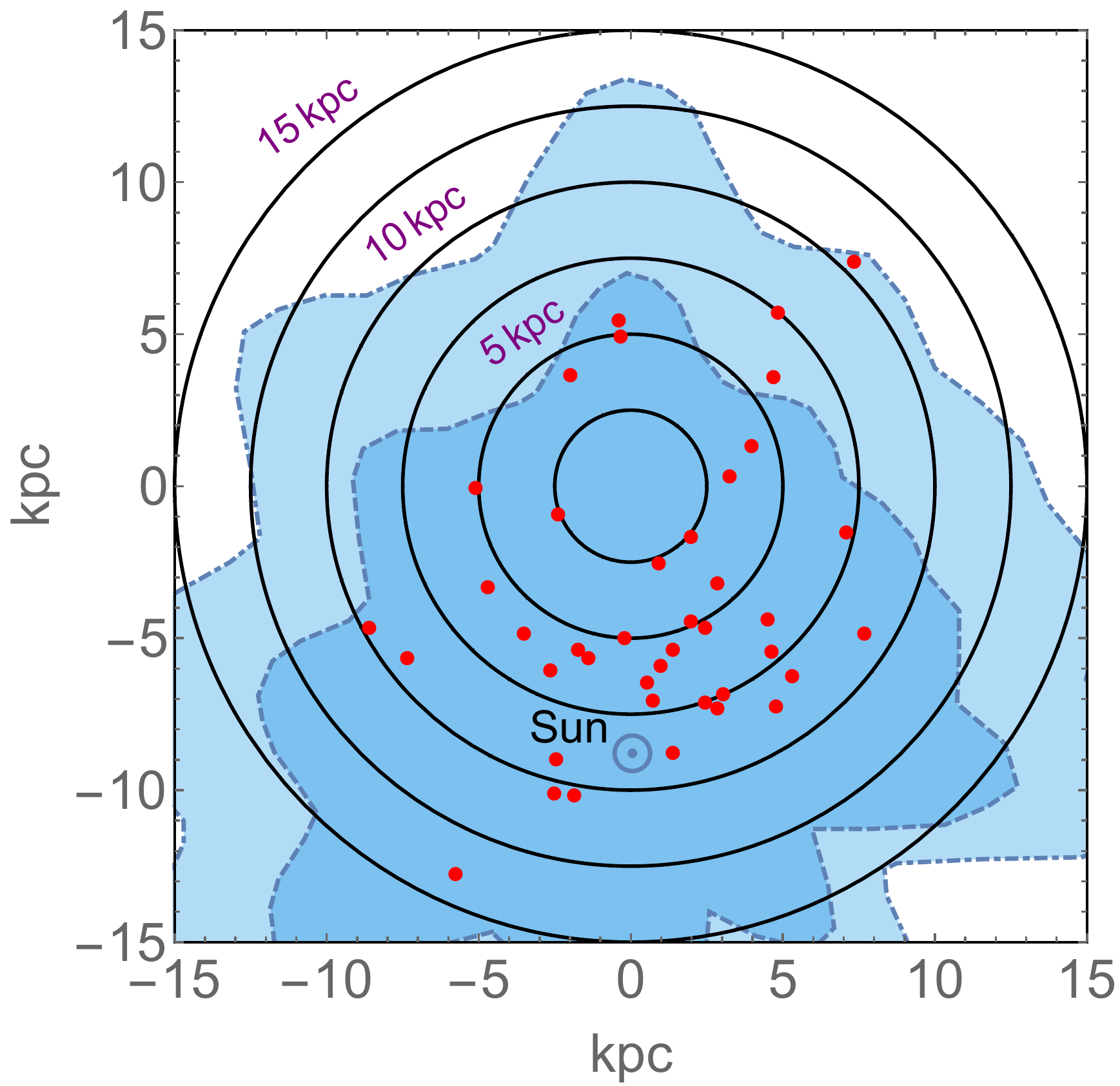}{0.99}
\caption{Illustrative view of the surface density of HMXBs in the
  Galaxy. The red points indicate positions of HMXBs. The dot-dashed
  and dashed curves show the regions of the Galaxy, within which the
  INTEGRAL Galactic survey detects all sources with luminosities $>
  10^{35.5}~{\rm erg} {\rm s}^{-1}$ and $> 10^{35}~{\rm erg} \
  s^{-1}$.}
\label{fig:density}
\end{figure}

The most recent catalogues show 114 HMXBs~\cite{Liu:2007tu} and
about 130 low-mass X-ray binaries (LMXBs)~\cite{Liu:2007ts}. The INTEGRAL/IBIS
nine-year Galactic plane survey, limited to $|b| < 17^\circ$, contains
82 high-mass and 108 low-mass sources~\cite{Krivonos:2012sd}.  The
sensitivity of this survey is about $10^{-11}~{\rm erg} \ {\rm s}^{-1}
\ {\rm cm}^{-2}$ in the 17-60 keV energy band, which ensures detection
of sources with luminosities $\agt 10^{35}~{\rm erg} \ {\rm s}^{-1}$
within half of the Galaxy ($\alt 9$~{\rm kpc} from the Sun) and $\agt
5 \times 10^{35}~{\rm erg} \ {\rm s}^{-1}$ over the entire Galaxy
($\alt 20~{\rm kpc}$ from the Sun); see Fig.~\ref{fig:density} . The number of X-ray binaries in
the Galaxy brighter than $2 \times 10^{34}~{\rm erg} \ {\rm s}^{-1}$
is thought to comprise 325 HMXBs and 380 LMXBs~\cite{Grimm:2001vd}.
These estimates may be uncertain by a factor of approximately two due to
our limited knowledge of the source spatial distribution, rendering them consistent with the observations from the
surveys reported above. Taken together this suggests an upper limit of
$\mu$QSOs in the Galaxy of ${\cal O} (100)$~\cite{Paredes:2003marti}.

About twenty $\mu$QSOs have been discovered so far. An illustrative
sample can be found in Table~\ref{microquasartable}.  Note that the
estimated jet luminosity of LS 5039 is relatively low, implying that
we can in principle use this source to estimate a lower bound on the
neutrino production efficiency required to be consistent with
observation. Note also that the only source with $L_{\rm jet}$ less than
that for LS 5039 has been observed in bursting and quiescent
states. In Table~\ref{microquasartable} we quote the quiescent value which is about a
factor of two lower than for the case of bursting state~\cite{Massi}.  

\begin{table*}
\caption{Properties of $\mu$QSOs in the Galaxy. \label{microquasartable}}
\begin{tabular}{ccccccc}
  \hline
  \hline
  ~~~Classification~~~ & ~~~Name~~~ & position (J2000.0) & ~~~distance [kpc]~~~ &
  ~~~$L_{\rm jet}$ [erg/s]~~~  & ~~~Reference~~~ \\
  \hline 
HMXB & LS I +61 303 &  ~~~~~$(02^{\rm h}40^{\rm m}31.70^{\rm s},
+61^\circ13'45.6'')$~~~~~ & 2 & $5.69 \times 10^{36}$ &\cite{Distefano:2002qw} \\
HMXB & CI Cam &  $(04^{\rm h}19^{\rm m}42.20^{\rm s},
+55^\circ59'58.0'')$ & 1 &  $5.66 \times 10^{37}$
&\cite{Distefano:2002qw} \\
LMXB & GRO J0422+32 & $ (04^{\rm h}21^{\rm m}42.70^{\rm s},
+32^\circ54'27.0'')$   & 3  & $4.35 \times 10^{37}$ &  \cite{Distefano:2002qw} \\ 
  LMXB & XTE J1118+480 & $(11^{\rm h}18^{\rm m}10.79^{\rm s},
    +48^\circ02'12.3'')$ & 1.9 & $3.49 \times 10^{37}$ &
    \cite{Distefano:2002qw} \\
    LMXB  & GS 1354-64 & $(13^{\rm h}58^{\rm m}09.70^{\rm s},
    -64^\circ44'05.0'')$ & 10 & $3.62 \times 10^{37}$ &  \cite{Distefano:2002qw} \\
    LMXB & Circinus X-1 &  $(15^{\rm h}20^{\rm m}40.84^{\rm s},
    -57^\circ10'00.5'')$ & 10  & $7.61 \times
    10^{38}$ & \cite{Distefano:2002qw} \\
    LMXB & XTE J1550-564 & $(15^{\rm h}50^{\rm m}58.67^{\rm s},
    -56^\circ28'35.3'')$ &  2.5 & $2.01 \times
    10^{38}$ & \cite{Distefano:2002qw} \\
    LMXB & Scorpius X-1 & $(16^{\rm h}19^{\rm m}55.09^{\rm s},
    -15^\circ38'24.9'')$ & 2.8 & $1.04 \times
    10^{38}$ & \cite{Distefano:2002qw} \\
LMXB & GRO J1655-40 & $(16^{\rm h}54^{\rm m}00.16^{\rm s},
    -39^\circ50'44.7'')$ & 3.1 & $1.6 \times 10^{40}$ & \cite{Distefano:2002qw}\\
LMXB & GX 339-4 &  $(17^{\rm h}02^{\rm m}49.40^{\rm s},
    -48^\circ47'23.3'')$ & 8 & $3.86 \times 10^{38}$ & \cite{Distefano:2002qw,Zdziarski:2004hv} \\
LMXB & 1E 1740.7-2942 & $(17^{\rm h} 43^{\rm m}54.82^{\rm s},
    -29^\circ44'42.8'')$ & 8.5 & $10^{36} - 10^{37}$ & \cite{BoschRamon:2006fw} \\
LMXB & XTE J1748-288 & $(17^{\rm h}48^{\rm m}05.06^{\rm s},
    -28^\circ28'25.8'')$ & 8 & $1.84 \times 10^{39}$ & \cite{Distefano:2002qw}\\
LMXB & GRS 1758-258 & $(18^{\rm h}01^{\rm m}12.40^{\rm s},
    -25^\circ44'36.1'')$ & 8.5  & $10^{36} - 10^{37}$ & \cite{Soria:2011dn}\\
HMXB & V4641 Sgr & $(18^{\rm h}19^{\rm m}21.63^{\rm s},
  -25^\circ24'25.9'')$ & 9.6 & $1.17 
  \times 10^{40}$ & \cite{Distefano:2002qw}   \\
  HMXB& LS 5039 & $(18^{\rm h}26^{\rm m}15.06^{\rm s},
  -14^\circ50'54.3'')$ & 2.9 & $8.73  \times 10^{36}$ & \cite{Distefano:2002qw} \\
    HMXB & SS 433 & $(19^{\rm h}11^{\rm m}49.57^{\rm s},
    +04^\circ58'57.8'')$ &4.8 & $1.00 \times 10^{39}$ & \cite{Distefano:2002qw} \\
LMXB & GRS 1915+105 & $(19^{\rm h}15^{\rm m}11.55^{\rm s},
    +10^\circ56'44.8'')$ & 12.5 & $2.45 \times 10^{40}$ &
    \cite{Distefano:2002qw} \\  
 HMXB & Cygnus X-1 & $(19^{\rm h}58^{\rm m}21.68^{\rm s}, +35^\circ12'05.8'')$
    & 2.1& $10^{36} - 10^{37}$ & \cite{Russell:2007ig} \\
    HMXB & Cygnus X-3 & $(20^{\rm h}32^{\rm m}25.77^{\rm s}, +40^\circ57'28.0'')$
    & 10 & $1.17 \times 10^{39}$ & \cite{Distefano:2002qw} \\
      \hline
    \hline
\end{tabular}
\end{table*}

\begin{figure}[tbp]
\postscript{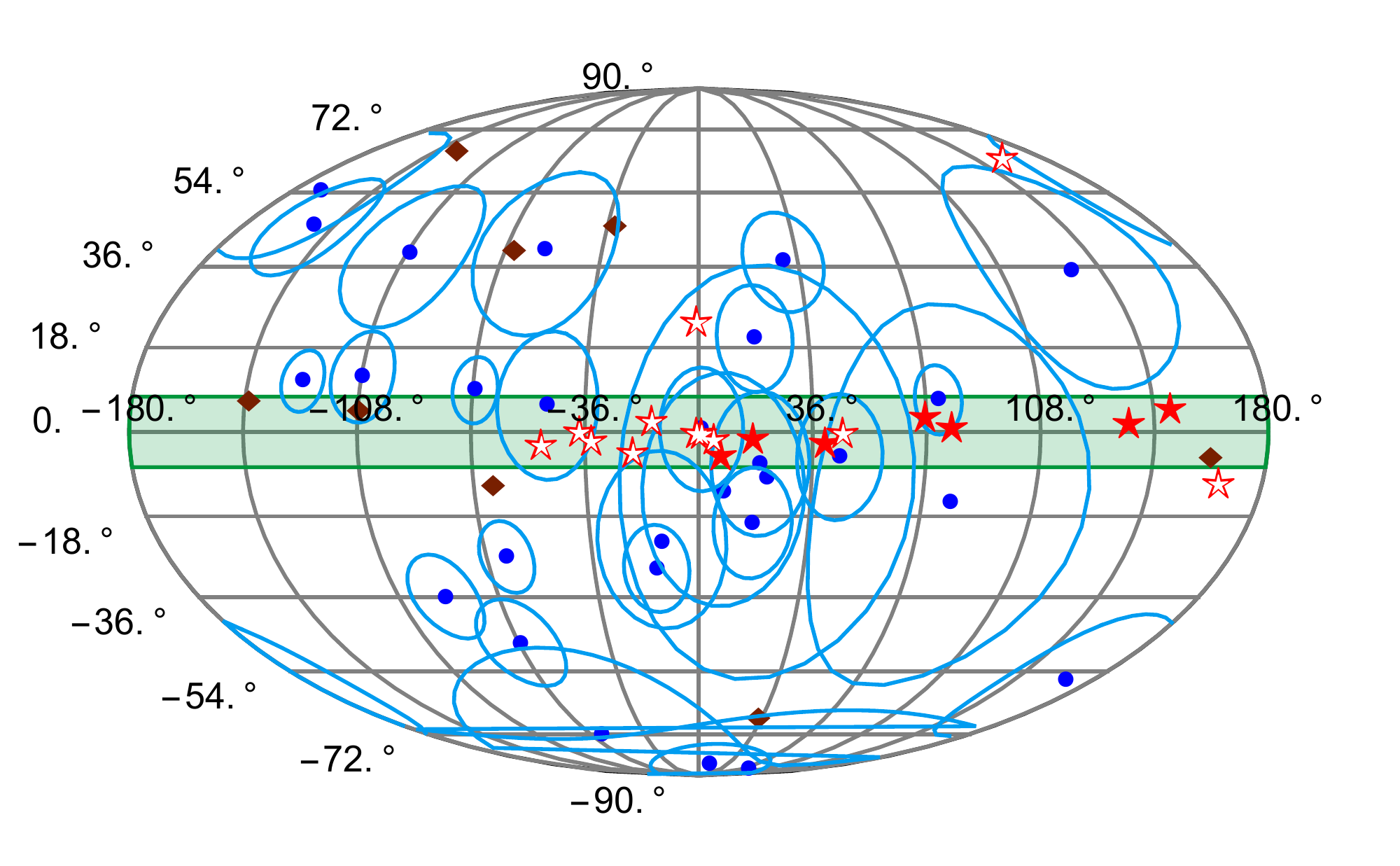}{0.99}
\caption{Comparison of IceCube event locations~\cite{Aartsen:2013bka}
  with Galactic $\mu$QSOs in a Mollweide projection. The 27 shower
  events are indicated by circles and the 8 track events by diamonds. The
  solid stars indicate the 7 $\mu$QSOs classified as HMXB and the
  outlined 
  stars the 12 $\mu$QSOs classified as LMXB.  The shaded band delimits
  the Galactic plane.}
\label{fig:skymap-mQSO}
\end{figure}

A comparison among all IceCube events and the Galactic $\mu$QSO
population is shown in Fig.~\ref{fig:skymap-mQSO}. Not surprisingly given
the size of the localization error, the two PeV neutrino events with
arrival direction consistent with the Galactic plane can be associated
with $\mu$QSOs within 1$\sigma$ uncertainties.

It appears that the impulse from supernovae explosions can eject a
system from its original position in the disk into the halo. In fact a
number of $\mu$QSOs have been observed with very high
velocities. For instance, XTE J1118-480 moves at $200~{\rm km} \ {\rm s}^{-1}$ in an
eccentric orbit around the Galactic Center~\cite{Mirabel:2001ay}. Additionally, the position
and velocity of Scorpius X-1 suggest it is a halo object~\cite{Mirabel:2003zr}.  Such speedy
objects are called runaway $\mu$QSOs. LS 5039 qualifies as a such
runaway $\mu$QSO with a velocity of $150~{\rm km} \ {\rm s}^{-1}$. Its computed
trajectory suggest it could reach a galactic latitude of $\sim
12^{\circ}$. The IceCube analysis search for multiple correlation in
the Galactic plane favors latitudes less than about $\pm 7.5^\circ$, which is not
inconsistent with the latitude reached by runaway
$\mu$QSOs.

The next to highest energy neutrino event is not in the Galactic
plane. It is also interesting to note that the position of this PeV
event is within 10 degrees in the hottest spot of IceCube search~\cite{Aartsen:2012gka} for
PeV $\gamma$-ray sources~\cite{Ahlers:2013xia}. If it turns out that PeV photons and
neutrinos are generated at the same sites, then observation of
coincidences implies these sites must be within the Galaxy, given the
short mean free path of PeV photons, which is less than
10~kpc. Conceivably, this could be associated with an as-yet
undiscovered $\mu$QSO.

At about $2~{\rm kpc}$ from Earth, there is another HMXB system with
similar characteristics to LS 5039.  LS I +61 303 has been detected at
all frequencies, including TeV and GeV energies~\cite{Albert:2006vk}.
Observations of persistent jet-like features in the radio domain at
$\sim 100~{\rm mas}$ scales prompted a classification of the source as
a $\mu$QSO~\cite{Massi:2003kf}, but subsequent observations at $\sim 1
- 10~{\rm mas}$ scales, covering a whole orbital period, revealed a
rotating elongated feature that was interpreted as the interaction
between a pulsar wind and the stellar wind~\cite{Dubus:2006ze}. More
recently, evidence favoring LS I +61 303 as the source of a very short
X-ray burst led to the analysis of a third alternative: a magnetar
binary~\cite{Bednarek:2009ur}. This binary system has also been
suspected to be a high energy neutrino
emitter~\cite{Torres:2006ub}. The source has been periodically
monitored by the AMANDA and IceCube
collaborations~\cite{Ackermann:2004aga}. The most recent analysis
leads to a 90\% CL upper limit on the neutrino flux at the level
$E_\nu^2 \Phi_{90} (E_\nu) = 1.95 \times 10^{-9}~{\rm GeV} \, {\rm
  cm}^{-2} \, {\rm s}^{-1}$~\cite{Aartsen:2014cva}. This implies that
if we were to consider LS 5039 as a standard neutrino source of the
$\mu$QSO population then $\gamma$'s and $\nu$'s should be produced
well above the base of the jet, without $\gamma$-ray absorption. For
such a case, the predicted neutrino flux is compatible with an
independent analysis presented in~\cite{Razzaque:2013uoa}, which
assumes the neutrino cluster arrives from the direction of the
Galactic center. Such a flux is also compatible with studies described
in~\cite{Anchordoqui:2013dnh}, which also postulate a Galactic center
origin, but with steeper spectral indices.  Finally, we stress that
the predicted high energy neutrino flux that can typify the $\mu$QSO
population is about an order of magnitude below the 90\% upper limit
reported by the ANTARES Collaboration~\cite{Adrian-Martinez:2014wzf},
see Fig.~\ref{fig:tres}.

In summary, if we assume the luminosity of LS 5039 truly typifies the
power of a $\mu$QSO then we should adopt as fiducial $L_\nu^{^{LS\, 5039}}  \approx
10^{33}~{\rm erg} \ {\rm s}^{-1}$, otherwise we will be inconsistent
with the IceCube limit on LS I +61 303. However, it is important to
stress that the value of $L_\nu^{^{LS\, 5039}}$ we will adopt to typify the population is very
conservative for far away sources, as one can observe in
Table~\ref{limits}. In closing, we note that though the IceCube bounds
are currently the most stringent, ANTARES has the potential to
discover exceptionally bright bursting sources in the Southern sky~\cite{Adrian-Martinez:2014ito}.

\begin{table}
\caption{90\% C.L. upper limits on the squared energy weighted flux of $\nu_\mu + \nu_{\bar
    \mu}$  in units of $10^{-9}~{\rm GeV} \ {\rm cm}^{-2}
  \ {\rm s}^{-1}$. \label{limits}}
\begin{tabular}{cccc}
\hline
\hline
Name & ~~~~~$E_\nu^2 \Phi_{^{90\% {\rm C.L.}}}^{_{\rm IceCube}}$~~~~~ &
~~~~~$E_\nu^2 \Phi^{_{\rm ANTARES}}_{^{90\% {\rm C.L.}}}$~~~~~& Reference\\
\hline
LS I 63 303 & 1.95 & $-$ & \cite{Aartsen:2014cva} \\
Circinus X-1 & $-$ & 16.2 & \cite{Adrian-Martinez:2014wzf}\\
GX 339-4 & $-$ & 15.0 & \cite{Adrian-Martinez:2014wzf} \\
LS 5039 & $-$ & 19.6 &\cite{Adrian-Martinez:2014wzf}\\ 
SS 433 & 0.65  & 23.2 & \cite{Aartsen:2014cva,Adrian-Martinez:2014wzf} \\
Cygnus X-3 & 1.70 & $-$ & \cite{Aartsen:2014cva}\\
Cygnus X-1 & 2.33 & $-$ & \cite{Aartsen:2014cva} \\
\hline
\hline
\end{tabular}
\end{table}

\section{High energy Neutrinos from Galactic microquasars }
\label{sec-tres}

Galactic $\mu$QSOs have long been suspected to be sources of high
energy neutrinos~\cite{Levinson:2001as}. In this section, we consider
the overall contribution of these candidate sources to the diffuse
neutrino flux, assuming LS 5039 is the nearest source and typifies the
$\mu$QSO population. We improve the procedure sketched
elsewhere~\cite{Anchordoqui:2013dnh}, in which the Earth was assumed
to be at the edge of the Galactic disk. In our current approach we
place the Earth in its actual position (about $8~{\rm kpc}$ from
the Galactic center) and perform the requisite integrations
numerically.  We further enhanced our previous analysis by considering
several source distributions. Firstly, we assume the sources are uniformly
distributed. Secondly, we assume the source density decreases
exponentially with distance from the Galactic center. These extremes are likely to
bound the true source distribution. Finally, we consider a more
realistic distribution to describe the particular case of $\mu$QSOs.

\begin{figure}
\postscript{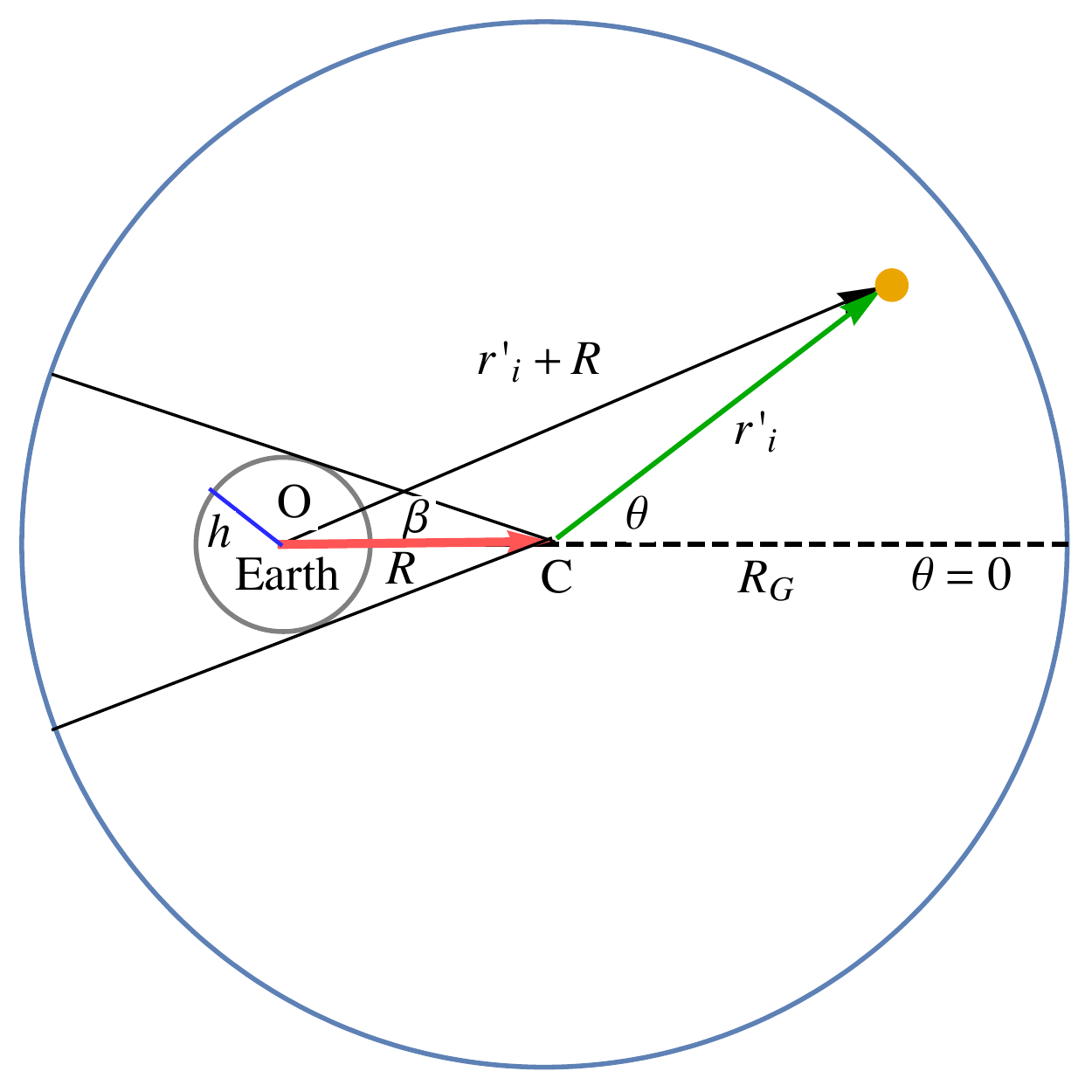}{0.8}
\caption{Sketch used to arrive at Eqs.~(\ref{cali1}) and (\ref{cali2}). Notice that we take
  account of the approximate location of the Earth in the Galactic
  disk. $h$ is a void  placed around the Earth to regularized
  the integration (see text). \label{left-right}}
\end{figure}

The ensuing discussion will be framed in the context of the thin disk
approximation. We model the Milky Way as a cylinder of radius $R_G =
15~{\rm kpc}$ and thickness $\delta = 1~{\rm kpc}$.  Consider the
situation displayed in Fig.~\ref{left-right} in which the observer $O$
is 
at the Earth, located at a distance $R = 8.3~{\rm kpc}$ from the center
of the Galaxy $C$. Denote the vector from $O$ to $C$ by $\vec R$, from
$C$ to the source $S_i$ by $\vec r_i^{\, \prime}$ and from $O$ to $S_i$ by $\vec
r_i$; then $ \vec r_i =\vec{R} + {\vec r_{i}}^{\,\prime}$ and so $r_{i}^{2} =
 R^{2} + r_{i}'^{2} + 2Rr_{i}' \, \cos\theta$. The integrated energy  weighted
total neutrino flux from the {\it isotropic} Galactic source distribution with normal
 incidence at $O$ is
 \begin{eqnarray}
4\pi \int_{E_1}^{E_2} \! \! E_\nu  \Phi(E_\nu)  dE_\nu & = & \frac{1}{4 \pi} \sum_{i}
        \dfrac{L_{\nu,i}}{r_{i}^{2}} \nonumber \\
 & = & \frac{1}{4 \pi} \sum_i \frac{L_{\nu,i}}{R^2 + 2 R r' \cos \theta
   + {r'}^2},
\label{sofia}
\end{eqnarray}
where $L_{\nu,i}$ is the power output of source $i$ and $\theta$ is
the angle subtended by $\vec r_i^{\, \prime}$ and $\vec R$.  Assuming
equal power for all sources, $L_{\nu,i} = L_\nu^{^{{\rm LS}\; 5039}}$,
we convert the sum to an integral
	\begin{eqnarray}
	4\pi \int_{E_1}^{E_2} \! \! E_\nu  \Phi(E_\nu)  dE_\nu  & = &
        \frac{ L_\nu^{^{{\rm LS} \; 5039}}}{4 \pi}  \nonumber \\
 & \times & \iint \dfrac{ \sigma (r') \,  r'dr'd\theta
        }{R^{2}+r'^{2}+2Rr' \cos\theta} \,, 
\label{nueve}
	\end{eqnarray} 
where $\sigma (r')$ is the source number density.  Any infrared divergence in
        (\ref{nueve}) is avoided by
        cutting off the integral within the void of radius $h$ as
        shown in Fig.~\ref{left-right}. For the sector of the
        circle {\it (i)} containing the observer, the integral in
        (\ref{nueve}) can be written as 
\begin{eqnarray}
{\cal I}_1 & = & 
 \int_{\pi + \phi}^{\pi - \phi}d\theta \int_{0}^{r_1 }
          \dfrac{ \sigma(r') \ r'dr'  }{R^{2}+r'^{2}+2Rr' \cos \theta} \nonumber \\
&+ &\int_{\pi + \phi}^{\pi -\phi}d\theta \int_{r_2}^{R_G }
          \dfrac{ \sigma(r') \ r'dr' }{R^{2}+r'^{2}+2Rr'cos \theta }  ,
\label{cali1}	
\end{eqnarray} 
where $\sin \phi = h/R$.  To determine $r_1$ we use the cosine law,
	$h^{2} = r_1^2 + R^{2} - 2Rr_1 \cos \beta $,
\begin{equation}
 r_1 = R \cos \beta \pm \sqrt{h^{2}- R^{2}  \sin^{2} \beta },
\label{once}
	\end{equation}
        where $\beta = \pi - \theta$.  For $\beta = 0$, we must
        recover $r_1 = R -h$ and so we take the minus sign in
        (\ref{once}).  The geometry of the problem then allows
        identification of $r_2$ as the solution with the positive sign
        in (\ref{once}).
      For the sector of the
        circle  {\it (ii)} outside the observer, the integral in (\ref{nueve})
        becomes
\begin{equation}
{\cal I}_2 = \int_{0}^{R_G } \int_{-\pi + \phi}^{\pi - \phi}
          \dfrac{ \sigma (r') \ r' dr' d\theta}{R^{2}+r'^{2}+2Rr' \cos \theta} \, .
\label{cali2}
\end{equation}
Putting all this together, for
$E_1 \sim 100~{\rm TeV}$ and $E_2 \sim 1~{\rm PeV}$, the diffuse neutrino flux on Earth is given by
\begin{eqnarray}
E_\nu^2 \ \Phi (E_\nu) &=& \frac{  d^2 E_\nu^2  \phi_\nu(E_\nu)  }{ 4 \pi}   \, \left({\cal I}_1 + {\cal I}_2 \right) \nonumber \\
&=& \frac{  d^2 \zeta }{ 4 \pi}   \, \left({\cal I}_1 + {\cal I}_2 \right) \nonumber \\
&=& \frac{  L_\nu^{^{{\rm LS}\; 5039}}   }{ 16 \pi^2 \ln 10}   \, \left({\cal I}_1 + {\cal I}_2 \right) \ .
\label{catorce}
\end{eqnarray} 
For $100~{\rm TeV} \alt E_\nu \alt 3~{\rm PeV}$, the IceCube Collaboration reports a flux 
\begin{equation}
\Phi (E_\nu) = 1.5 \times 10^{-8}
\left(\frac{E_\nu}{100~{\rm TeV}}\right)^{-2.15\pm 0.15}~({\rm GeV} \;
  {\rm cm}^{2} \; {\rm s} \;
{\rm sr})^{-1} \,, \nonumber
\label{IceCube-flux}
\end{equation}
assuming an isotropic source distribution and
democratic  flavor ratios~\cite{Aartsen:2013bka}.  For
direct comparison with IceCube data, (\ref{catorce}) can be rewritten
in standard units using the fiducial value of the source
luminosity derived in the previous section,
\begin{eqnarray}
E_\nu^2 \ \Phi (E_\nu)  \approx   1.27 \times 10^{-9} \ {\rm GeV \
  cm^{-2} \ s^{-1} \ sr^{-1}}  \frac{{\cal I}_1 + {\cal I}_2}{{\rm kpc}^{\rm 2}}  \ .
\end{eqnarray} 
The integrals ${\cal I}_1$ and $ {\cal I}_2$ have been computed
numerically for various void configurations assuming equal power
density per unit area of the disk, that is $\sigma_\Theta (r') =N/ \pi
R_G^2$, where $N$ is the total number of sources. The results are
given in Table~\ref{tabla1}.  The number of sources required to
provide a dominant contribution to IceCube data depends somewhat on
the size of the void $h$. For $h \approx 3~{\rm kpc}$, about 900
sources are needed to match IceCube observations. This corresponds to
a total power in neutrinos of about $6 \times 10^{36}~{\rm erg} \ {\rm
  s}^{-1}$. If we assume that these accelerators also produce a hard
spectrum of protons with equal energy per logarithmic interval, then
the estimate of the total power needed to maintain the steady observed
cosmic ray flux is more than two orders of magnitude
larger~\cite{Anchordoqui:2013qsi,Gaisser:1994yf}.

\begin{table}[htb]
\caption{Results for numerical integration of   (\ref{cali1}) and (\ref{cali2}), assuming
                 various source distributions, and equivalent point
                 source number $N$. The values listed in the table
                 are in units of kpc$^{-2}$. \label{tabla1}} 
\begin{tabular}{cccc}
\hline\hline
~~~$h$ [kpc]~~~ & ~~~$({\cal I}_1 + {\cal I}_2)_{\Theta} $~~~ & ~~~$({\cal I}_1 + {\cal I}_2)_{\rm exp}$~~~ & ~~~$({\cal I}_1 + {\cal I}_2)_{\mu {\rm QSO}}$~~~ \\
\hline
1 & 0.0224 N & 0.0211 N & 0.0273 N \\
2 & 0.0163 N & 0.0178 N & 0.0193 N \\
3 & 0.0127 N & 0.0163 N & 0.0146 N \\
4 & 0.0101 N & 0.0154 N & 0.0113 N \\
5 & 0.0081 N & 0.0148 N & 0.0088 N \\ 
\hline
\hline
\end{tabular}
\end{table}

\begin{table}
\caption{Number of sources required for each distribution to dominate
  the neutrino flux reported by the IceCube Collaboration.  \label{tabla3}}
\begin{tabular}{cccc}
\hline\hline
~~~~~~~$h$ [kpc]~~~~~~~ & ~~~~~~~$N_\Theta$~~~~~~  & ~~~~~~~$N_{\rm exp} $~~~~~~~ & ~~~~~~~$N_{\mu{\rm QSO}}$~~~~~~~ \\
\hline
1 & $ 527$ & $560$ & $433$  \\
2 & $ 724$ & $663$ & $612$  \\
3 & $ 930$ & $725$ & $809$  \\
4 & $ 1169$ & $767$ & $1045$  \\
5 & $ 1458$ & $ 798$ & $1342$  \\ 
\hline
\hline
\end{tabular}
\end{table}

In this note we have advocated a scenario in which a nearby source
contributes significantly to the overall flux, rendering it
anisotropic.  Should this be the case, {\it the isotropic contribution
  to the overall flux must be smaller than that derived based on the
  assumption that all IceCube events contribute to the isotropic
  flux.} To model the isotropic background of the nearby source
scenario we  duplicate the
procedure substituting in (\ref{nueve}) an exponential distribution of
sources which is peaked at the Galactic center, $ \sigma_{\rm exp} (r')=
n_{0} \, e^{-r'/r_{0}}$.  We normalize the distribution to the total
number of sources in the Galaxy, $ N = \int_{0}^{2\pi} d\theta
\int_{0}^{ R_{G} } \sigma_{\rm exp}(r') r' dr'$. Because we have two parameters
we need an additional constraint.  We choose to restrict the
percentage of the total number of sources beyond the distance $R-h$ to
the galactic edge $R_G$,
\begin{equation}
P_{R-h} = 2 \pi \int_{R-h}^{R_G} n_0 e^{ -r'/r_0 } \ r' dr'   \ ,
\end{equation}
We choose to take $P_{R-h} = 10 \%$.  The number of sources required
to produce a diffuse neutrino flux at the level reported by the
IceCube Collaboration is given in Table~\ref{tabla3}, for different
values of $h$.

Recent studies~\cite{Lutovinov:2013ga,Grimm:2001vd} of persistent
HMXBs in the Milky Way, obtained from the deep INTEGRAL Galactic plane
survey~\cite{Krivonos:2012sd}, provide us a new insight into the
population of $\mu$QSOs. The HMXB surface densities (averaged
over corresponding annuli) are given in Table~\ref{tabla2}. It can be
seen that the overall distribution of surface density in the Galaxy
has a peak at galactocentric radii of $5 - 8~{\rm kpc}$ and that HMXBs
tend to avoid the inner $2 - 4~{\rm kpc}$ of the
Galaxy~\cite{Grimm:2001vd}. Therefore, it is clear that a simple
exponential disk component is not a good description for the radial
distribution. In the spirit of~\cite{Dehnen:1996fa}, we assumed a
source density distribution in the form
\begin{equation}
\sigma_{\mu {\rm QSO}}  (r') = N_0 \exp \left[-\frac{R_0}{r'} -
  \frac{r'}{R_0} \right]   \,,
\label{sigmamuqso}
\end{equation}
where the first term in the exponential allows for the central density
depression. To describe the observed central depression for high-mass
X-ray binaries we take $R_0= 4~{\rm kpc}$~\cite{Grimm:2001vd}. This
 is also supported by a fit to the data in Table~\ref{tabla2}. The number of sources required
to produce a diffuse neutrino flux at the level reported by the
IceCube Collaboration is given in Table~\ref{tabla3}, for different
values of $h$. For a void of 1~kpc, which is the distance to the
nearest source in Table~\ref{microquasartable} (CI Cam), about 500
sources are needed to reproduce IceCube observations.

\begin{table}
\label{lfpars}
\caption{Best fit parameters of the HMXB spatial density
  distribution. \label{tabla2}}
\begin{tabular}{cc}
\hline
\hline
~~~~~~~~~~$r'$~[kpc]~~~~~~~~~~ &~~~~~~~~~~$N(L>10^{35}~{\rm erg} \, {\rm s}^{-1})~{\rm kpc}^{-2}$~~~~~~~~~~\\
\hline
0-2      & $0.0\pm0.05$(syst.)\\[1mm]
2-5      & $0.11^{+0.05}_{-0.04}$(stat.)$\pm0.02$(syst.)\\[1mm]
5-8      & $0.13^{+0.04}_{-0.03}$(stat.)$\pm0.01$(syst.)\\[1mm]
8-11    & $(3.8^{+2.1}_{-1.2})\times10^{-2}$(stat.)$\pm6.5\times10^{-3}$(syst.)\\[1mm]
11-14  & $(6.2^{+7.2}_{-4.3})\times10^{-3}$(stat.)$\pm4.8\times10^{-3}$(syst.)\\[1mm]
\hline
\hline
\end{tabular}
\end{table}

It is worth commenting on an aspect of this analysis which may seem
discrepant at first blush. We find that some 500 $\mu$QSOs are
required to satisfy energetics requirements, while current
catalogs/estimates describe about 100 such known objects. This is not
so worrying for the following reasons. First, we have considered only
the lower bound on $\mu$QSO jet luminosity, which may vary by up to
three orders of magnitude in the catalog listings (see
Table~\ref{microquasartable}). In this sense our estimated required
number of $\mu$QSOs that can plausibly explain the IceCube data is a
conservative one.  Secondly, when considering the nearby source
scenario we did not re-evaluate the background conditions, which would
yield a smaller isotropic flux.\footnote{Evaluating the background, of
  course, require detailed knowledge of detector properties and
  properly belongs to the territory of the IceCube Collaboration.}
Again, this is a conservative path. Thus, the analysis presented
herein adheres to a ``cautious'' approach throughout, lessening (or
eliminating) concerns about the discrepancy between our estimates of
the required number of $\mu$QSOs versus the cataloged quantities. We
then conclude that $\mu$QSOs could provide the dominant contribution
to the diffuse neutrino flux recently observed by IceCube.

\section{Constraints from gamma rays and baryonic cosmic rays}
\label{cuatro-extra}

\begin{figure}[tbp]
\postscript{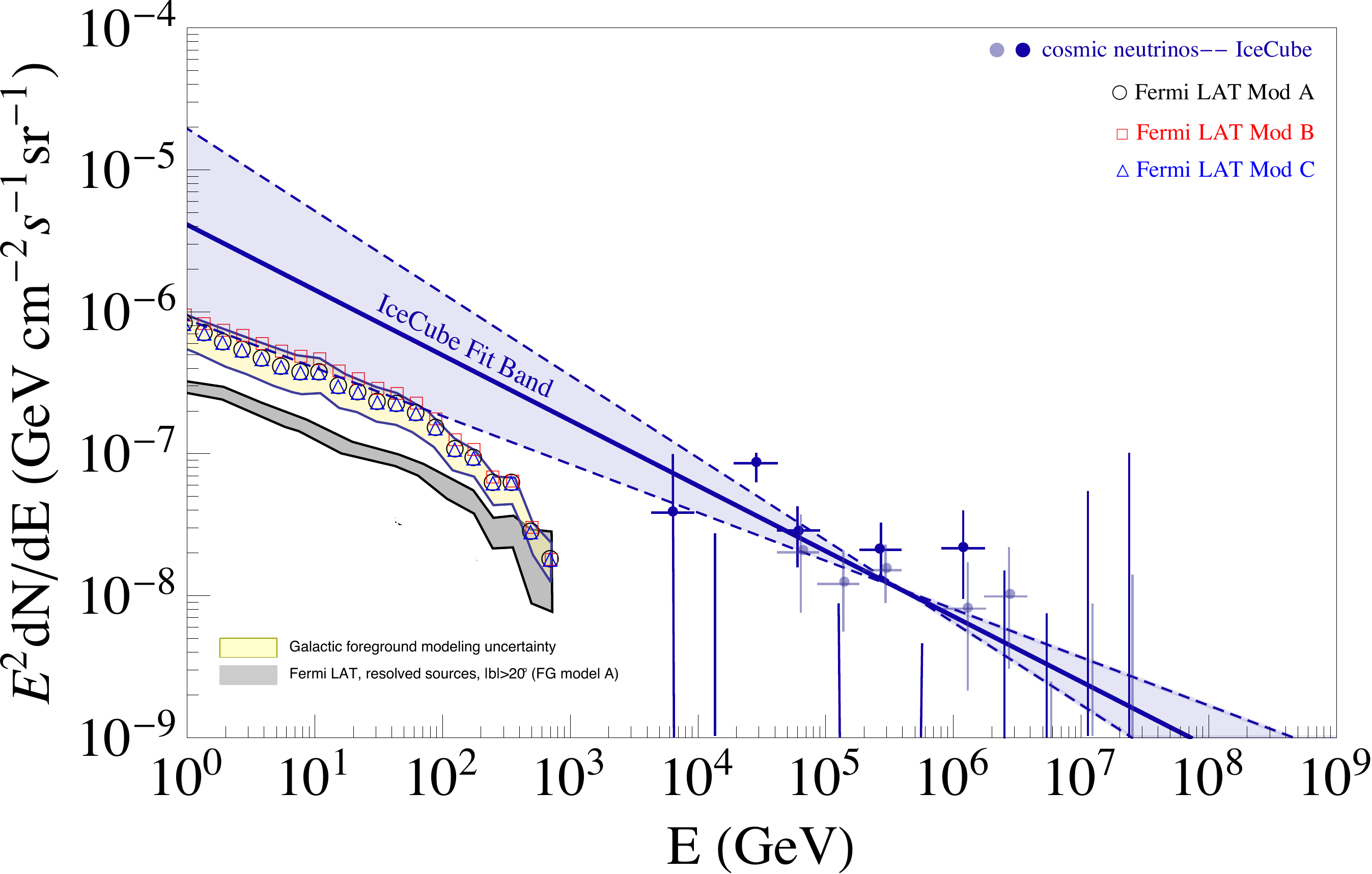}{0.99}
\caption{The open symbols represent the total extragalactic
  $\gamma$-ray background for different foreground (FG) models as
  reported by the Fermi Collaboration~\cite{Ackermann:2014usa}. For details on the modeling of 
the diffuse Galactic foreground emission in the benchmark FG models A,
B and C, see~\cite{Ackermann:2014usa}. The cumulative intensity from
  resolved Fermi LAT sources at latitudes $|b| > 20^\circ$ is indicated by a (grey)
  band. The solid symbols indicate the neutrino flux reported by
  the IceCube Collaboration. The best fit to the data (extrapolated
  down to lower energies), $\Phi (E_\nu) = 2.06^{+0.4}_{-0.3} \times
  10^{-18} (E_\nu/10^5~{\rm GeV})^{-2.46 \pm 0.12}~{\rm GeV}^{-1}\ {\rm
      cm}^{-2} \ {\rm s}^{-1} \ {\rm sr}^{-1}$, is also shown
  for comparison. \label{fig:fermi+icecube}}
\end{figure}

Very recently the IceCube Collaboration has extended their neutrino
sensitivity to lower energies~\cite{Aartsen:2014muf}.  One intriguing
result of this new analysis is that the spectral index which best fits
the data has steepened from $2.15 \pm 0.15$ to $2.46 \pm 0.12$.  If one
assumes the neutrino spectrum follows a single power law up to about
10~GeV, then the latest data from the Fermi
telescope~\cite{Ackermann:2014usa} can be used to constrain the
spectral index assuming the $\gamma$-rays produced by the $\pi^0$'s
accompanying the $\pi^\pm$'s escape the source. In such a scenario,
Fig.~\ref{fig:fermi+icecube} shows that only a relatively hard
extragalactic spectrum is consistent with the data.  On the other
hand, the Galactic photon flux in the 10~GeV region is about an order
of magnitude larger than than the extragalactic flux; this allows
easier accommodation of a softer single power law spectrum.  For the
Galactic hypothesis, however, one must consider an important caveat,
namely that the expected photon flux in the PeV range has been
elusive~\cite{Borione:1997fy}. However, a recent refined analysis of archival data from the
EAS-MSU experiment~\cite{Fomin:2014ura} has confirmed previous claims
of photons in the 10~PeV region.  This analysis also results in a
larger systematic uncertainty at all energies, relaxing previously
reported bounds in the PeV range. While previous bounds were
marginally consistent with non-observation of PeV photons expected to
accompany the IceCube neutrinos~\cite{Anchordoqui:2013qsi}, this new
less stringent bound is more comfortably consistent.

There is an additional interesting consequence of the new IceCube
data. The neutrino spectral index should follow the source spectrum of
the parent cosmic rays. We have shown
elsewhere~\cite{Anchordoqui:2013qsi,Anchordoqui:2014hia} that a
spectral index of $\sim 2.4$ is required for consistency with current
bounds on cosmic ray anisotropy.  Further credence regarding our
best-fit spectral index has been recently developed via numerical
simulations~\cite{Giacinti:2014xya}. It is worth stressing that our
discussion regarding source energetics assumes the canonical Fermi
index of $\alpha = 2$.  Given the current level of uncertainties on the
atmospheric neutrino background, the spatial distribution and total
number of microquasars, as well as the large variation in microquasar
jet luminosities (see Table~\ref{microquasartable}), shifting our assumed spectral
index from $\alpha = 2$ to $\alpha = 2.4$ will have little impact on
the arguments concerning energetics explored herein. In the future,
improved measurements all-round will require a considerably more
elaborate analysis, including detailed numerical simulations.

\section{Conclusions}
\label{sec-cuatro}

Motivated by recent IceCube observations we have re-examined the idea
that $\mu$QSOs are high energy neutrino emitters. We considered the
particular case of LS 5039, which as of today represents the source
with lowest p-value in the IceCube sample of selected
targets~\cite{Aartsen:2013bka}. We have shown that if LS 5039 has a
compact object powering jets, it could accelerate protons up to above
about 30 PeV. These highly relativistic protons could subsequently
interact with the plasma producing a neutrino beam that could reach
the maximum observed energies, $E_\nu \agt~{\rm PeV}$. There are two
extreme possibilities for neutrino production: {\it (i)} close to the
base of the jet and {\it (ii)} at the termination point of the jet. By
normalizing the accompanying photon flux to H.E.S.S. observations in
the TeV energy range~\cite{Aharonian:2005eb,Aharonian:2006vu} we have shown that, for the first scenario,
photon absorption on the radiation field leads to a neutrino flux
${\cal O} (10^{-8} E_\nu^{-2}~{\rm GeV}^{-1} \, {\rm cm}^{-2} \, {\rm
  s}^{-1})$. Should this be the case, the neutrino flux almost saturates
the current upper limit reported by the ANTARES Collaboration~\cite{Adrian-Martinez:2014wzf}. The
second possibility yields a flux of neutrinos which is about an order
of magnitude smaller. {\it A priori} these two extreme flux predictions are
partially 
consistent with existing data.  However, one can ask why a source with
similar characteristics (LS I +61 303) which is in the peak of the
field of view of IceCube has not been already discovered. The current
90\% CL upper limit on LS I +61 303 reported by the IceCube
Collaboration is ${\cal O} (10^{-9} E_\nu^{-2}~{\rm GeV}^{-1} \, {\rm
  cm}^{-2} \, {\rm s}^{-1})$, favoring neutrino production near the
end of LS 5039 jets.

We have also generalized our discussion to the population of $\mu$QSOs
in the Galaxy. Using the spatial density distribution of high-mass
X-ray binaries obtained from the deep INTEGRAL Galactic plane survey
and assuming LS 5039 typifies the $\mu$QSO population we have
demonstrated that these powerful compact sources could provide the
dominant contribution to the diffuse cosmic neutrino flux. Of course,
a complete picture which accommodates all the shower events outside
the galactic plane may well require an extragalactic component.
Indeed most of the istropic background is dominated by muon tracks.
Explaining the possible isotropy of shower events may eventually prove
only to be possible by considering extragalactic sources. Future
IceCube observations will test the LS 5039 hypothesis, providing the
final verdict for the ideas discussed in this paper.

\section*{Acknowledgements}
This work was supported in part by the US NSF grants CAREER
PHY-1053663 (LAA, LHMdaS), PHY-1314774 (HG), and PHY-1205854 (TCP), the NASA
grant NNX13AH52G (LAA, TCP), AYA-ESP2012-39115-C03,
AYA-ESP2013-47816-C4, and MULTIDARK Consolider CSD2009-00064
(BJV). LAA and TCP thank the Center for Cosmology and Particle
Physics at New York University for their hospitality.

\end{document}